\documentclass[preprint,showpacs,preprintnumbers,amsmath,amssymb]{revtex4}

\usepackage{graphicx}
\usepackage{dcolumn}
\usepackage{bm}

\begin{document}


\title{A general approach to statistical modeling of physical laws:\\nonparametric regression}

\author{Igor Grabec}
\altaffiliation[Also at ]{Amanova,
Kantetova 75, 1001 Ljubljana, Slovenia.}

\affiliation{Faculty of Mechanical Engineering, University of Ljubljana,\\
A\v{s}ker\v{c}eva 6, PP 394, 1001 Ljubljana, Slovenia}
\email{igor.grabec@fs.uni-lj.si}
\homepage{http://www.fs.uni-lj.si/lasin/}


\date{\today}

\begin{abstract}
Statistical modeling of experimental physical laws is based on the probability density function of measured variables. It is expressed by experimental data via a kernel estimator. The kernel is determined objectively by the scattering of data during calibration of experimental setup. A physical law, which relates measured variables, is optimally extracted from experimental data by the conditional average estimator. It is derived directly from the kernel estimator and corresponds to a general nonparametric regression. The proposed method is demonstrated by the modeling of a return map of noisy chaotic data. In this example, the nonparametric regression is used to predict a future value of chaotic time series from the present one. The mean predictor error is used in the definition of predictor quality, while the redundancy is expressed by the mean square distance between data points. Both statistics are used in a new definition of predictor cost function. From the minimum of the predictor cost function, a proper number of data in the model is estimated.

\end{abstract}



\maketitle

\section{\label{sec:1}Introduction}

A basic task of physical description of natural phenomena is to express relations between experimental data about measured variables in terms of physical laws \cite{fe}. Since the corresponding analytical modeling essentially depends on the intuition of the explorer performing it, an ambiguity surrounds this basic task and there thus arises a question how this could be avoided. This problem becomes of fundamental practical importance when developing intelligent electronic systems for automatic modeling of physical laws \cite{gs}. The ambiguity could be avoided if a unique objective method of modeling was found that would take into account common properties of experimental observations and of transitions from experimental data to models. The aim of this article is to show how such a method could be developed from basic principles of probability and statistics, as well as to demonstrate an example of its applicability. 

A common property of all experimental explorations is that each experiment corresponds to a process proceeding from preparation to execution. If we want a selected experiment to yield any information about the phenomenon under observation, then the result of the experiment may not be determined in advance i.e. several outcomes of the experiment must be possible. The next common property is repeatability of experiments. Consequently, a correct presentation of experimental observations requires the use of a distribution of experimental results and this must be related to the concept of probability. The probability distribution is, therefore, a common basis for the description of natural properties in terms of experimental data \cite{re}, while the transition from experimental data to an analytical expression of the corresponding probability distribution function is the crucial problem of modeling. An objective solution of this problem represents statistical modeling of the probability distribution function by a nonparametric kernel estimator if the kernel is determined by a calibration of the experimental setup \cite{ig,ig2,ig3}. For this purpose, the central theorem of probability theory and the maximum entropy principle provide a quite general route to the specification of the kernel function of the estimator. In this case, an experimental physical law, which represents a relation between observed variables, can also be generally expressed by applying the theory of optimal statistical estimators. The resulting nonparametric regression is the conditional average (CA), which can be automatically extracted from the probability density function (PDF) of experimental data in a measurement system. The complete approach to modeling thus appears objective, independent of the intuition of the observer and, consequently, generally applicable for automatic execution. Due to these convenient properties, CA is widely applicable in various fields of natural and technical sciences \cite{gs}.
 
A nonparametric expression of the PDF by the kernel estimator has already been proposed by Parzen \cite{par,dh}, but weaknesses of his proposal are that the kernel function is arbitrarily introduced, and that there is an assumption that its width should decrease to zero when the number of data is increased to infinity. In order to avoid this weakness, we specify the kernel function objectively by the scattering of the measurement system output during calibration \cite{par,dh}. The only ambiguity in the expression of the PDF is then related to the number of experimental data, which according to Parzen's assumption should not be limited. Since an infinite number of experiments cannot be performed, there arises a fundamental question: "How many experiments is it reasonable to perform in order to explore the phenomenon properly by a given experimental setup?" Intuitively, we can conclude that it is reasonable to repeat experiments for as long as they bring new information. However, with an increasing number of experiments, the acquired data points become ever more concentrated in the sample space and consequently the repetition of the experiments becomes redundant. This is observed when distances between data points become comparable to the width of the kernel function. This reasoning led recently to a specification of an information cost function $C$ 
\cite{ig,ig2,ig3,les,ris,ris2,ct,kol}. For this purpose the indeterminacy of measurements was first expressed in terms of information entropy, which further led to definition of the experimental information $I$ and the redundancy $R$ of experiments. Using these statistics, the information cost function was expressed by the difference $C=R-I$. From the position of its minimum, a proper number of experiments can then be objectively determined \cite{ig,ig2,ig3}.

Estimation of the information cost function is related to the calculation of integrals, which is inconvenient in a multivariate case. Therefore, another statistic, with similar properties but more simple calculation, is sought. Since it has been shown previously that the predictor quality exhibits similar properties to the experimental information, we utilize it here in the definition of the predictor cost function. From its minimum, a proper number of experiments can also be estimated. If this is used as a proper number for the adaptation of the nonparametric regression to data provided by experiments, the modeling of the corresponding physical law can be performed automatically on a data acquisition system of the experimental setup. To demonstrate this possibility, we first briefly describe the nonparametric regression and then turn to the definition of the predictor quality, redundancy and cost function. Properties of all statistics are subsequently demonstrated in the modeling of a return map corresponding to a noisy chaotic process. 

\section{\label{sec:2} Fundamentals of nonparametric modeling}
\subsection{\label{sec:2a}Description of kernel function}

Let us consider a phenomenon that can be described by just two joint variables, since the generalization to a multivariate case is straightforward. A single result of joint measurement is represented by the couple ${\bf z}=(x,y)$. We next assume that the phenomenon can be characterized statistically by repetition of measurements yielding sample points ${\bf z}_n =(x_n ,y_n )$ in the joint span of a two channel instrument $S_{\bf z}=S_x \otimes S_y$. 
 
Since the instruments are generally subject to stochastic disturbances, the results of measurements are scattered even during repetition of calibration \cite{les}. The scattering can be described by the data provided by a series of repeated simultaneous calibrations of both instrument channels. For this purpose, we have to perform a joint measurement on an object representing two physical units $u_x$ and $u_y$ which we denote together by the joint unit ${\bf u}=(u_x,u_y)$. The scattering of instrument outputs during calibration is characterized by the joint PDF $\psi({\bf z}|{\bf u})$, which we call the scattering function (SF) \cite{ig,gs,les}. When the interaction between both channels is negligible, the SF is given by the product $\psi({\bf z}|{\bf u})=\psi(x|u_x)\psi(y|u_y)$. Without loss of generality, we further consider a case with equal channels which are subject to mutually independent random disturbances that do not depend on ${\bf u}$. In such cases, the central limit theorem of probability theory, as well as the maximum entropy principle, suggest that we  express the SF of a particular channel by the Gaussian function:
\begin{equation}
{\rm g}(x-u_x,\sigma)\,=
\frac{1}{\sqrt{2\pi}\,\sigma}\exp \biggl[-\frac{(x-u_x)^2}{2\sigma}\biggr]
\end{equation}
The parameters $u_x$, $\sigma$ represent the mean value and standard deviation of signal $x$ at the calibration and can be statistically estimated from given data. The joint SF is then determined 
by the product $\psi({\bf z}-{\bf u})={\rm g}(x-u_x,\sigma)\,{\rm g}(y-u_y,\sigma)$. 

When reporting experimental results, experimentalists most often only specify  mean values and standard deviations of variables during calibration. The maximum entropy principle tells us that, in such cases, the Gaussian function is the best choice for SF \cite{gs,les}.

\subsection{\label{sec:2b} Nonparametric estimation of PDF pertaining to experimental data}

When we perform a single measurement, we get a sample ${\bf z}_1=(x_1,y_1)$ that represents the mean value of ${\bf z}$ during measurement and, therefore, we express the PDF as $\psi({\bf z}-{\bf z}_1)=\psi(x-x_1)\psi(y-y_1)$. When we repeat the measurements $N$ times, we get a set of samples $\{{\bf z}_i ,\,1\le i \le N\}$, by which we model the joint PDF by the statistical average:
\begin{equation}
f({\bf z})\,=\,\frac{1}{N}\,\sum_{i=1}^N \psi ({\bf z}-{\bf z}_i) 
\label{pdfxy}
\end{equation}
that represents the kernel estimator.

Properties of the particular components $x,y$ are described by the marginal PDFs $f(x), f(y)$. They are obtained from the joint PDF by integration with respect to one component, for example:
\begin{equation}
f(x)\,=\,\int_{S_y} f({\bf z}) dy\,=\,\frac{1}{N}\,\sum_{i=1}^N \psi (x-x_i).
\label{pdfxe}
\end{equation}
For modeling natural laws, the most important is the conditional PDF of the variable $y$ at a given value of $x$, defined as: 
\begin{equation}\label{cpdfe}
f(y|x)\,=\, \frac{f({\bf z})}{f(x)}\,=\,\frac{\sum_{i=1}^N \psi ({\bf z}-{\bf z}_i) }{\sum_{j=1}^N \psi (x-x_j) }
\end{equation}

\subsection{\label{sec:2c}Estimation of a physical law}

\begin{figure}
\centering
\includegraphics[width=2.5in]{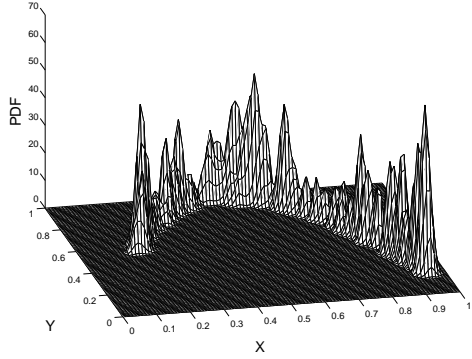}
\caption{\label{figpdfz} The joint PDF $f({\bf z})$ utilized to demonstrate the properties of the conditional average estimator.}
\end{figure}
Distributions of joint experimental data, for example that shown in Fig.\,\ref{figpdfz}, often resemble a ridge along some hypothetical line $y_{\rm o}(x)$, which we want to extract from the given data in an optimal way. For this purpose, we select from a set of joint data only those that pertain to some selected $x$. These joint data generally exhibit various values of $y$ which we try to represent by a single value called the predictor of the variable $y$ from a given value $x$. We consider as an optimal predictor of the hypothetical $y_{\rm o}$ the value $y_p$ at which the mean square prediction error is minimal:
\begin{equation}
{\rm E} [(y_p - y)^2|x]\, = \,{\rm min}(y_p).
\end{equation}
Here ${\rm E} [\ldots|x]$ denotes the operation of statistical averaging at given condition 
$x$. The minimum satisfies the equation: $d {\rm E} [(y_p - y)^2|x]/dy_p=0$ that yields as the optimal predictor $y_p$ the conditional average: 
\begin{equation}\label{CA}
y_p(x)\,=\,{\rm E} [y|x]\,=\,\int_{S_y} y \,f(y|x) \,dy 
\end{equation}
By using Eq.\,(\ref{cpdfe}), we obtain for the conditional average the expansion:
\begin{equation}\label{CAN}
y_p(x)\,=\,\frac{\sum_{i=1}^N y_i \psi (x-x_i,\sigma)}{\sum_{j=1}^N \psi (x-x_j,\sigma)}=\sum_{i=1}^N y_i B_i (x).
\end{equation}
The coefficients of this expansion are sample values $y_i$, while the basis functions are
\begin{equation}\label{C}
B_i (x)\,=\,\frac{\psi (x-x_i,\sigma)}{\sum_{j=1}^N \psi (x-x_j,\sigma)},
\end{equation}
and satisfy the following conditions:
\begin{equation}
\sum_{i=1}^N B_i (x)=1\quad ,\quad 0 \leq B_i (x) \leq 1. 
\end{equation}
The basis functions $B_i (x)$ can be interpreted as a normalized measure of similarity between the given value of $x$ and its sample value $x_i$. At a given $x$, the sample value $y_m$ contributes most to the estimated value $y_p(x)$ whose complementary sample value $x_m$ is most similar to $x$.

The calculation of $y_p(x)$ corresponds to an associative recall of memorized items, which is a property of an intelligence. Therefore, the estimator $y_p(x)$ could be treated as a basis for the development of a machine intelligence based on modeling of natural laws. The conditional average given in Eq.\,\ref{CAN} in fact corresponds to a normalized radial basis function neural network which is equivalent to a multilayer perceptron -- the basic paradigm used in the theory of artificial neural networks \cite{gs,ha}. 

\section{\label{sec:3} Characteristics of the model}
\subsection{\label{sec:3a} Predictor quality}

A predictor maps the stochastic variable $x$ to a new stochastic variable $ y_p$ that generally differs from the variable $y$. When the variables $x,y$ are related by some hypothetical physical law $y_{\rm o}(x)$ and the measurement noise is small, the first and second statistical moments ${\rm E}[ y- y_p]$, ${\rm E}[ (y- y_p)^2]$ of the prediction error are also small. The second moment is: ${\rm E}[ (y- y_p)^2]={\rm Var} (y)+{\rm Var} ( y_p)-2{\rm Cov} (y, y_p)+[{\rm m}(y)-{\rm m}( y_p)]^2$, where ${\rm E}, {\rm m}, {\rm Var},{\rm Cov}$ denote statistical average, mean value, variance and covariance respectively. In the case of statistically independent variables $y$ and $ y_p$ with equal mean values, the last two terms are zero and we get: ${\rm E}[ (y- y_p)^2]={\rm Var} (y)+{\rm Var} ( y_p)$. With respect to this relation, we define the predictor quality relatively by the formula
\begin{eqnarray}
Q&=&1-\frac{{\rm E}[ (y- y_p)^2]}{{\rm Var} (y)+{\rm Var} ( y_p)} \nonumber \\
&=&\frac{2{\rm Cov} (y, y_p)}{{\rm Var} (y)+{\rm Var} ( y_p)}-\frac{[{\rm m}(y)-{\rm m}( y_p)]^2}{{\rm Var} (y)+{\rm Var} ( y_p)}
\label{Q}
\end{eqnarray}
The quality is $1$ if the prediction is exact: $y_p= y$, while it is $0$ if $y$ and $ y_p$ are statistically independent and have equal mean values. The quality $Q$ may be negative if ${\rm m}(y)\ne {\rm m}( y_p)$. For the predictor defined by the conditional average $y_p(x)\,=\,\int y \,f(y|x) \,dy$, we analytically obtain the equalities:
${\rm m}(y)={\rm m}(y_p)$ and ${\rm Cov} (y, y_p)= {\rm Var} ( y_p)$, which yield  
\begin{equation}
Q=\frac{2{\rm Var} ( y_p)}{{\rm Var} (y)+{\rm Var} ( y_p)}.
\label{QCA} 
\end{equation}
From the definition of the conditional average, it follows $0\le {\rm Var} ( y_p)\le {\rm Var}(y)$ and therefore $0\le Q\le 1$. This inequality need not be fulfilled exactly if CA is statistically estimated from a finite number of samples. 

With an increasing $N$, we generally expect that the CA statistically estimated by Eq.\,(\ref{CAN}) increasingly better represents the governing physical law and, consequently, that the corresponding predictor quality $Q$ on average increases to a certain limit value. As mentioned previously, an unlimited increase in the number of experiments is experimentally impossible and, consequently, there arises the question how to determine a proper number $N_{\rm o}$ of data that will yield a judicious estimation of the governing law. 

\subsection{\label{sec:3b} Redundancy and predictor cost function}

To answer the last question, we have analyzed various experimental cases which have shown us that, with an increasing number of experimental samples, the value of predictor quality generally stabilizes when the distance between data points becomes similar to the width $\sigma$ of the scattering function. Therefore, it is not reasonable to surpass significantly the corresponding number of data. This can be achieved if a ratio of  $\sigma$ and a proper measure of distance $\delta$ between neighbor data points is considered. For this purpose, we introduce $\delta$ over the mean value of minimum square distance between data points: $\delta^2={\rm E}[{\rm min}\{(x_i-x_j)^2+(y_i-y_j)^2)\};i=1\ldots N,j=1\ldots N,]$, and define a measure of redundancy of data by the relative variable:
\begin{equation}
R=2N\frac{\sigma^2}{\delta^2}
\label{Redund} 
\end{equation}
Since $\delta^2$ is comprised of two terms denoting contributions from $x$ and $y$ components, a factor $2$ is utilized in the nominator. The fraction $2\sigma^2/\delta^2$ represents an average increase of redundancy that is assigned to the acquisition of a new data point. In order to take into account acquisition of $N$ data points, factor $N$ is further used. With respect to this, we introduce the predictor cost function by the sum: 
\begin{eqnarray}
C&=&R - Q + 1 \nonumber\\
&=&2N \frac{{\sigma}^2}{\delta^2}+\frac{{\rm E}[ (y- y_p)^2]}{{\rm Var} (y)+{\rm Var} ( y_p)}.
\label{C} 
\end{eqnarray}
The constant $1$ is inserted in the first row in order to obtain a more simple expression in the second row of Eq.\,\ref{C}. In the same way as the definition of the information cost function given in \cite{ig2,ig3}, the cost function is here expressed in a relative form comprised of two terms: the first corresponds to the redundancy of experiments due to inaccurate measurements while the second represents the influence of acquisition of information about the phenomenon by experiments.  With an increasing number of samples $N$, the redundancy on average increases while the second term decreases with the decreasing error. Therefore, the cost function $C$ exhibits a minimum at some $N_{\rm o}$ that represents a proper number of data needed for the modeling of the physical law governing the phenomenon explored. However, the influence of the first term becomes prevailing when the distance between data points $\delta$ becomes essentially smaller than the width $\sigma$ of the scattering function.

\section{\label{sec:4} Example}

\begin{figure}
\centering 
\includegraphics[width=3.375in]{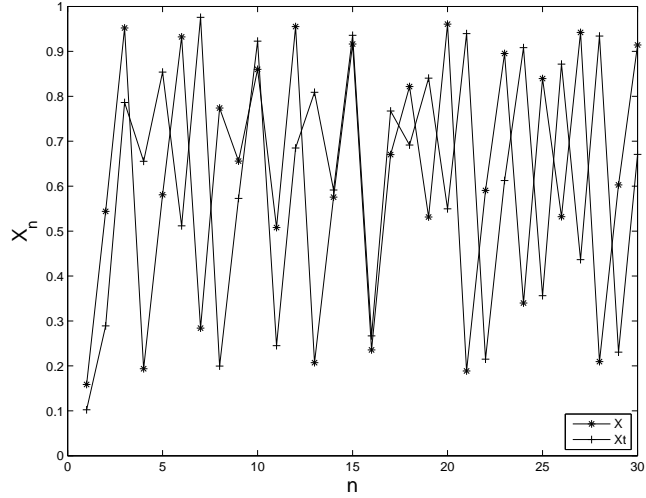}
\caption{\label{figxN} Records of the basic -- ($X$), and the testing -- ($Xt$) noise corrupted chaotic series.}
\end{figure}

To demonstrate the properties of the CA estimator, we utilize the data generated by a noise-corrupted chaotic return map with the span $S_x=(0,1)$. This example is used because similar cases often appear in the analysis of chaotic time series \cite{gs,mo}. The basic problem in such an analysis is to extract the return map from a given record of time series that is influenced by additive noise of instrumental origin. In our case, we apply analytically determined data to provide for a comparison between the original and extracted physical law and to make feasible an objective reproduction of the complete method. The basic governing law is here given by the logistic map:
\begin{equation}
\chi_{n+1}=3.8\, \chi_n (1-\chi_n),
\label{chagen}
\end{equation}
while the initial value $\chi_1$ is arbitrary selected from the interval $(0,1)$ using a random generator. To the values of generated chaotic series, the Gaussian noise $\nu$ of zero mean value and standard deviation $\sigma=0.1$ is added to simulate an additive noise of measurement. The iterative solution of Eq.\,\ref{chagen} then yields a series of noise corrupted chaotic values: $x_n=\chi_n + \nu_n$. Figure \ref{figxN} shows two records of such a series that were used in modeling and testing of the proposed method.

\begin{figure}
\centering 
\includegraphics[width=3.375in]{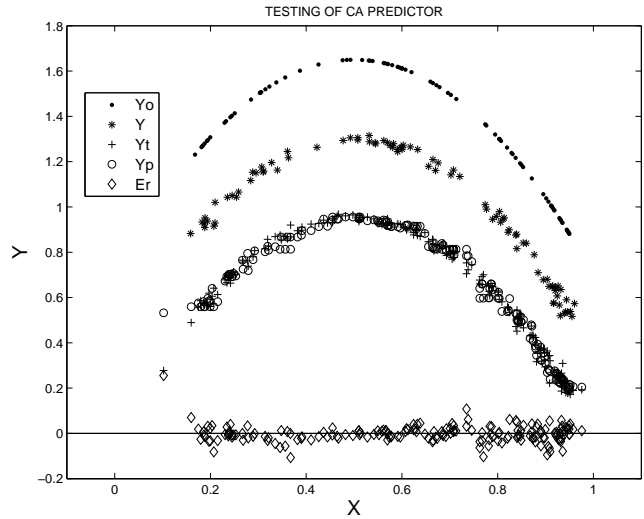}
\caption{\label{figCAN} Testing of the CA predictor. Graphs represent the governing law $y_{\rm o}$ and basic data $y$ -- (top two: $\cdot\cdot\cdot$ ; $\ast\ast\ast$), test $y_t$ and predicted data $y_p$ -- (middle two: $+++$ ; $\circ\circ\circ$ ), and prediction error $E_r=y_p-y_t$ -- (bottom: $\diamondsuit\diamondsuit\diamondsuit$). The upper two parabolas are displaced successively by 0.35 in the vertical direction for better visualization.}
\end{figure}

From the series $\{x_n\, ; n=1\ldots\}$, the joint samples of the basic variables $x,y$ are obtained by treating the successive value of $x_n$ as the dependent variable: $y_n=x_{n+1}$. The generator of data is thus analytically described by the rule:
\begin{eqnarray}
x_n&=&\chi_n + \nu_n \nonumber \\
y_n&=&x_{n+1},
\end{eqnarray}
while the governing law is given by $y_{\rm o}=3.8\, x(1-x)$. The sample points $\{x_n,y_n\,;\, n=1\ldots N\}$ are distributed along the corresponding parabola in the sample space. According to our previous treatment, the standard deviation $\sigma$ corresponds to the width of the instrument scattering function $\psi$. The joint PDF shown in Fig.\,\ref{figpdfz} is determined by the kernel estimator Eq.\,(\ref {pdfxy}) using $200$ data, while a reduced set of $30$ data is further utilized to demonstrate the properties of the conditional average estimator. The data obtained from the pure chaos generator are shown by $y_{\rm o} \cdot\cdot\cdot$ in the top parabola of Fig.\,\ref{figCAN}, while the basic noise-corrupted data $y \ast\ast\ast$ are shown by points scattered around pure data points. 

The conditional average estimator is obtained by inserting data from the basic data set into Eq.\,(\ref{CAN}). To demonstrate its performance, we additionally generated with different seeds of random generators a set 
of $N_t=60$ test data $\{ x_{t,i},y_{t,i} \}$. Based on data $x_{t,i}$ from this set, the corresponding values of $y_{p,i}$ are predicted by the CA estimator. The test and predicted data are shown in Fig.\,\ref{figCAN} by the middle two sets of points ($+++$ and $\circ\circ\circ$). The prediction error $Er=y_p-y_{t}$, calculated from both data sets, is presented by $\diamondsuit\diamondsuit\diamondsuit$ at the bottom of Fig.\,\ref{figCAN}. Relatively small differences between predicted and test points indicate that the properties of the governing law $y_{\rm o}(x)$ are properly modeled by the CA estimator. To confirm this qualitative conclusion, we next analyze the properties of statistics ${\rm E}[(y_e-y_t)^2], Q, \delta^2, R, C$ depending on the the number of data $N$ used in modeling. The number of test data is kept constant $N_t=60$ during calculation of these statistics. Properties of the statistical model of the governing law depend on sets of samples utilized in modeling and testing. To demonstrate this dependence, we repeated the modeling and testing three times using various statistical sample sets. 
\begin{figure}
\centering 
\includegraphics[width=3.375in]{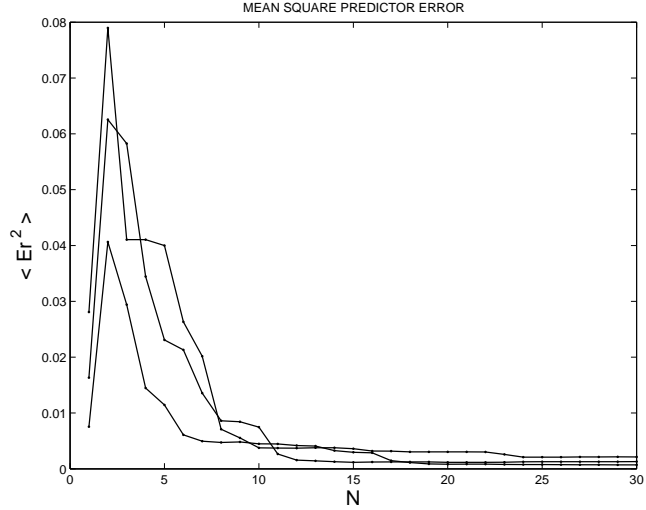}
\caption{\label{figEr} Mean square prediction error ${\rm E}[(y- y_p)^2]$ as a function of the data number $N$.}
\end{figure}

The mean square predictor error ${\rm E}[(y- y_p)^2]$ is presented in Fig.\,\ref{figEr} versus number of samples $N$. Its value varies statistically but, on average, it decreases with the increasing number $N$. Statistical fluctuations are largest at small $N$ and significantly depend on samples used in modeling. However, with the increasing $N$, the statistical fluctuations are ever less pronounced. If the number of test samples $N_t$ is much larger than the number of samples $N$, changing the testing sample set does not significantly influence the properties of estimated statistics, which is the case in our demonstration. This is the reason why we use the value $N_t=60$.
\begin{figure}
\centering 
\includegraphics[width=3.375in]{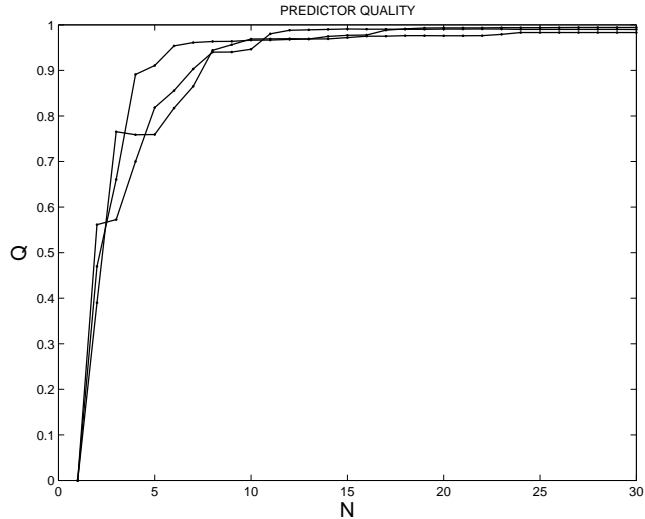}
\caption{\label{figQ} Predictor quality $Q$ as a function of the data number $N$.}
\end{figure}

The predictor quality $Q$, as determined from the prediction error, is presented in Fig.\,\ref{figQ} versus number of samples $N$. For each data set the statistical fluctuations decrease with increasing $N$ so that qualities calculated from different data sets converge to the same limit value. With increasing $N$, the curves determined from different data sets merge approximately at $N\sim 11$. The quality is there 
$\sim 0.97$ and rises to $\sim 0.98$ at $N=30$. At $N\sim 11$, the difference between the curves obtained from different data sets is about two orders of magnitude smaller than the corresponding quality. With respect to these properties, we could conjecture that in the present case about $11$ data values already provide for a judicious modeling of the governing law $y_{\rm o}(x)$ by the CA predictor. 

\begin{figure}
\centering 
\includegraphics[width=3.375in]{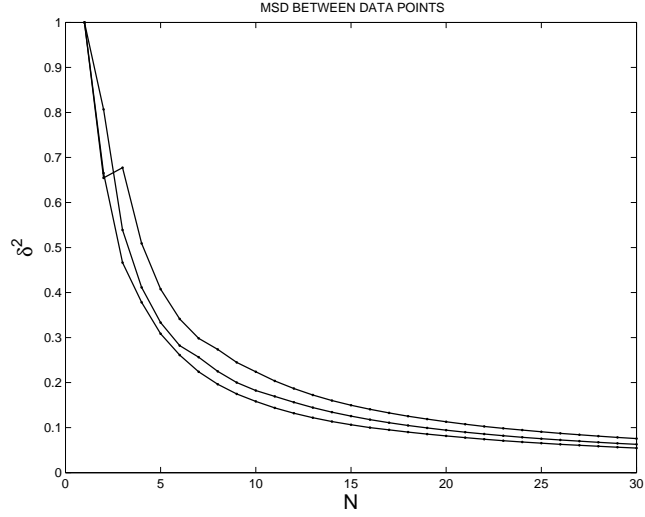}
\caption{\label{figDij2m} Mean square distance between data points $\delta^2$ as a function of the data number $N$.}
\end{figure}

\begin{figure}
\centering 
\includegraphics[width=3.375in]{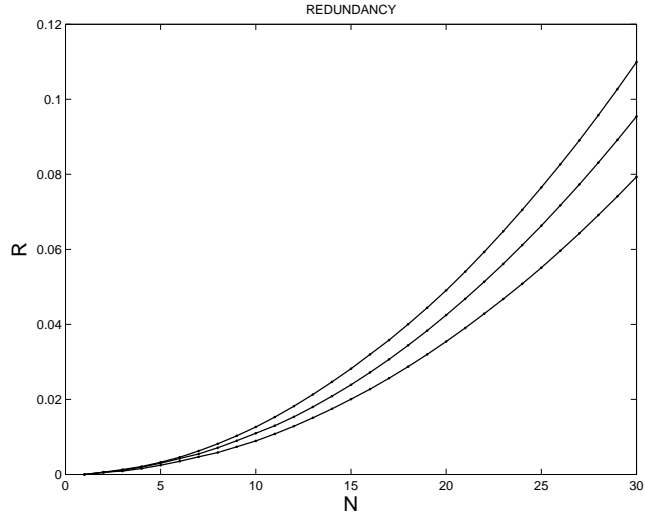}
\caption{\label{figR} Redundancy $R$ as a function of the data number $N$.}
\end{figure}

To confirm our last conjecture, we turn to the determination of the predictor cost function. For this purpose, let us first analyze the properties of the mean square distance between data points $\delta^2$. The corresponding graph, shown in Fig.\,\ref{figDij2m}, indicates that $\delta^2$ is rather monotonously decreasing with the number of samples with the approximate dependence being $\sim 1/N$. Consequently, the corresponding redundancy $R$ is increasing with $N$ similarly as $\sim N^2$. This conclusion is confirmed by the graph in Fig.\,\ref{figR}. 

\begin{figure}
\centering 
\includegraphics[width=3.375in]{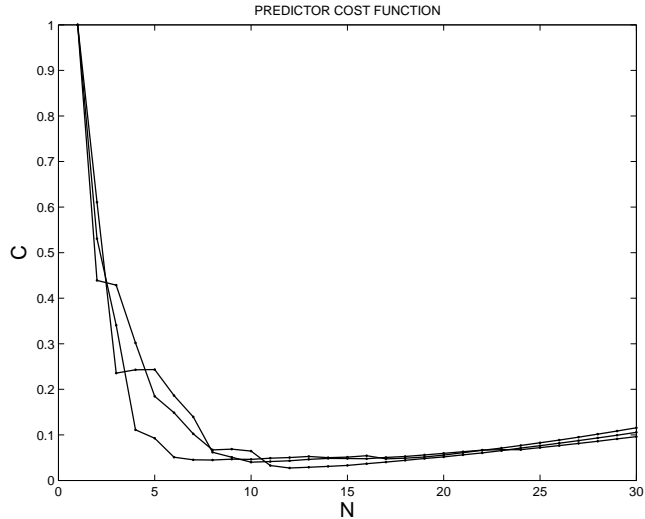}
\caption{\label{figC} Predictor cost function $C$ as a function of the data number $N$.}
\end{figure}

Following the definition given by Eq.\,\ref{C}, we obtain from the estimated error and the redundancy the predictor cost function $C$ shown in Fig.\,\ref{figC}. Its minimum is not very pronounced. From various statistical data sets, we obtain the estimates of the minimal value $C_{\rm o}=0.033\pm0.006$. The corresponding number $N_{\rm o}=10\pm 2$ confirms our previous conjecture stemming from the analysis of predictor quality. 

With an increasing number of samples $N$, the quality $Q(N)$ of the CA predictor exhibits a convergence to some limit value $Q_\infty$ that characterizes hypothetical maximum quality of proposed nonparametric statistical modeling. This limit value generally increases with the decreasing scattering width $\sigma$. Related to this, the minimal value of cost function is diminished and takes place at a larger $N_{\rm o}$\,; for instance at $\sigma=0.005$ we get $C_{\rm o}=0.018\pm0.003$ and $N_{\rm o}=14\pm 3$. However, the limit value of the quality $Q_\infty$ is less than $1$ if $1/\sigma$ and $N$ are finite. This means that it is not possible to exactly determine the governing physical law $y=y_{\rm o}(x)$ from joint data obtained by an instrument influenced by stochastic disturbances.\\

\section{\label{sec:5} Discussion} 

Our method of estimation of natural laws from given data can be simply generalized to multivariate cases by substituting corresponding vectors for the variables $x,y$. Such modeling has already been applied in a variety of examples stemming from physical \cite{mgg}, technical \cite{gs,tgp}, economic \cite{gs,tgp} and medical environments \cite{gs,grag,gfg}. Particularly in economic and medical environments, phenomena are often characterized by many variables that could be either informative or disturbing. Due to the complexity of such cases, there usually exists little or no information about a possible function that could describe the governing law. In relation to this, researchers are faced with the problem of how to define complexity and to reduce it by extracting informative variables from a given set \cite{be}. Alongside mutual information, the predictor quality could also be applied for this purpose. For instance, it has been recently shown in the field of medicine how an analysis of predictor quality can provide for an ordering of variables and the extraction of a set that yields an optimal predictor of the disease healing process \cite{grag,gfg}. Such an analysis makes feasible further progress towards the origins of the treated disease. 

The value of the proper number $N_{\rm o}$, as defined by the minimum of predictor cost function, could be interpreted as a measure of the complexity of an adequate predictor model. It is important that this measure depends only on the accuracy of observation and properties of the phenomenon represented by given experimental data.

In relation to the example demonstrated here, there emerges an important conclusion about the description of natural phenomena by physical laws in the form $y=y_{\rm o}(x)$. As long as such a law is considered as the only basis for the description of the phenomenon, it is not sufficient for a complete description, since no information is provided about the properties of the sample space of joint data. Consider a well known example -- the law $m=\rho V$ that relates the mass $m$, the volume $V$ and the density $\rho$ of an object. This law does not include the restriction $m\ge0$, and is in this aspect not complete. Similar, but much more complex, examples are met when treating chaotic phenomena and their strange attractors \cite{mo}. For example, the law applied here is a special case of the law $\chi_{n+1}=a\, \chi_n (1-\chi_n)$, with $a$ being a constant. Depending on the value of $a$ and the starting value $\chi_1$, the series $\{\chi_n\, ; n=1\ldots \}$ exhibits at large values of parameter $n\rightarrow\infty$ either a discrete or a continuous sample space. Moreover, in the continuous case, the sample space can be comprised of disconnected intervals which could hardly be predicted analytically. Similar, but still more cumbersome, is the situation if we consider chaotic processes with continuous parameters. Consequently, a governing law $y=y_{\rm o}(x)$ appears incomplete for description of the phenomenon. The most outstanding deficiency is that it does not include information about the structure of the sample space corresponding to the observed phenomenon. This deficiency does not appear if we consider as a basis for modeling the probability density function and estimate it nonparametrically, directly from measured joint data. The extraction of a law that describes a relation between variables can then be generally performed by using the conditional average estimator.  However, applications of simple parametrical laws, like $m=\rho V$, are of tremendous importance for analytical sciences and we do not expect that the proposed nonparametric models could substitute for them, although they are convenient for direct applications. Consequently, the question arises of how to find a univocal link between both paradigms of modeling.

\section{\label{sec:6} Conclusions}

Our approach indicates that the objectively introduced kernel estimator provides for a nonparametric statistical modeling of a quantitatively explored phenomenon. Since no a priori information about the form of the governing physical law is required, the modeling can be automatically performed by a computer in a measurement system. The proposed predictor cost function $C$ provides for estimating the proper number $N_{\rm o}$ of data needed for the modeling. Properties of the predictor cost function resemble those of information cost function \cite{ig2,ig3}, but its estimation is much more simple. The properties of the extracted model of the governing law can be quantitatively described by the predictor quality $Q$ and redundancy $R$ of data from which the governing law is extracted. This law represents the distribution of the variable $y$ at a given value $x$ by a single predicted value $y_p(x)$. Such a compressed representation generally corresponds to creation of information about the explored phenomenon \cite{ig2,ig3}. This is in contrast to the loss of information caused by stochastic disturbances in signal transmission channels \cite{sha}. If the extraction of information from observations is considered as a basis of natural intelligence \cite{ha,ka}, then a system capable of estimating a physical law from measured data autonomously must be treated as an intelligent unit. Such an interpretation provides a common basis for a unified treatment of experimental sciences and natural or artificial intelligence \cite{gs,ha,ka}.

\begin{acknowledgments}
This work was supported by The Ministry of Higher Education, Science and Technology of the Republic of Slovenia and EU -- COST.
\end{acknowledgments}


\begin{thebibliography}{}

\bibitem{fe} R. Feynman, {\em The Character of Physical Law} (The MIT Press,Cambridge, MA, 1994).
\bibitem{gs} I. Grabec and W. Sachse, {\em Synergetics of Measurement, Prediction and Control} (Springer-Verlag, Berlin, 1997).
\bibitem{re} R. E. Collins, Found. Physics {\bf 35}, 734 (2005). 
\bibitem{ig} I. Grabec, Eur. Phys. J. B {\bf 22}, 129 (2001).
\bibitem{ig2} I. Grabec, Eur. Phys. J. B {\bf 48}, 279 (2005), (DOI: 10.1140/epjb/e2005-00391-0).
\bibitem{ig3} I. Grabec, arXiv:cs.IT/0612027 {\bf v1 5}, (2006). 
\bibitem{par} E. Parzen, Ann. Math. Stat. {\bf 35}, 1065 (1962).
\bibitem{dh} R. O. Duda and P. E. Hart, {\em Pattern Classification and Scene Analysis} (J. Wiley and Sons, New York, 1973), Ch. 4.
\bibitem{les} J. C. G. Lesurf, {\em Information and Measurement} (Institute of Physics Publishing, Bristol, 2002).
\bibitem{ris} J. Risanen, {\em Complexity, Entropy, and the Physics of Information} (Addison-Wesley, 1990), ed. W. H. Zurek, 117-125.
\bibitem{ris2} J. Rissanen, IEEE Trans. Inf. Theory {\bf 42}, 40 (1996).
\bibitem{ct} T. M. Cover and J. A. Thomas, {\em Elements of Information Theory} (John Wiley \& Sons, New York, 1991).
\bibitem{kol} A. N. Kolmogorov, IRE Trans. Inf. Theory {\bf IT-2}, 102 (1956).
\bibitem{mo} F. C. Moon, {\em Chaotic and Fractal Dynamics} (John Wiley \& Sons, INC. New York, 1992).
\bibitem{mgg} S. Mandelj, I. Grabec and E. Govekar, Int. J. Bifurcation and Chaos {\bf 11}, 2731 (2001). 
\bibitem{tgp} M. Thaler, I. Grabec and A. Poredo\v{s}, Physica A {\bf 35}, 46 (2005).
\bibitem{grag} I. Grabec and D. Gro{\v s}elj, Comput. Methods in Biomech. Biomed. Engin. {\bf 6}, 319 (2003) 
\bibitem{gfg} I. Grabec, I. Ferkolj and D. Gro{\v s}elj, {\em Proc. of 2nd International Conference on Computational Intelligence in Medicine and Healthcare, Lisbon}, (CIMED-2005 Proceedings, ISBN: 0-86341-520-2,IEE, 2005), ed. J. M. Fonseca, 311-316 
\bibitem{be} C. H. Bennett, {\em Complexity, Entropy, and the Physics of Information} (Addison-Wesley, 1990), ed. W. H. Zurek, 137-148.
\bibitem{sha} C. E. Shannon and W. Weaver, {\em The Mathematical Theory of Communication} (Univ. of Illinois Press, Urbana, 1949).
\bibitem{ha} S. Haykin, {\em Neural Networks, A Comprehensive Foundation} (Mcmillan College Publishing Company, New York, 1994)
\bibitem{ka} D. J. C. MacKay {\it Information Theory, Inference, and Learning Algorithms}  (Cambridge University Press, Cambridge, UK, 2003)
 
\end{thebibliography}
\end{document}